\documentclass[pre,twocolumn,preprintnumbers,amsmath,amssymb,superscriptaddress]{revtex4-2}

\usepackage{color}
\usepackage[pdftex]{graphicx}
\usepackage{verbatim}
\usepackage{amssymb}
\usepackage{amsmath}
\usepackage{color}
\usepackage[export]{adjustbox}
\renewcommand{\figurename}{Fig.}

\usepackage{mathtools}

\usepackage{makecell}
\usepackage{enumerate}
\usepackage{mathtools}
\usepackage{stmaryrd}
\usepackage[toc,page]{appendix}
\usepackage{textcomp}
\usepackage{gensymb}
\usepackage[shortlabels]{enumitem}

\begin{document}

\title{Complete Visitation Statistics of 1d Random Walks}
\author{L\'eo R\'egnier}
\affiliation{Laboratoire de Physique Th\'eorique de la Mati\`ere Condens\'ee,
CNRS/Sorbonne University, 4 Place Jussieu, 75005 Paris, France}
\author{Maxim Dolgushev}
\affiliation{Laboratoire de Physique Th\'eorique de la Mati\`ere Condens\'ee,
CNRS/Sorbonne University, 4 Place Jussieu, 75005 Paris, France}
\author{S. Redner}
\affiliation{Santa Fe Institute, 1399 Hyde Park Road, Santa Fe, NM, USA 87501}
\author{Olivier B\'enichou}
\affiliation{Laboratoire de Physique Th\'eorique de la Mati\`ere Condens\'ee,
CNRS/Sorbonne University, 4 Place Jussieu, 75005 Paris, France}

\begin{abstract}
We develop a framework to determine the complete statistical behavior of a fundamental quantity in the theory of random walks, namely, the probability that $n_1, n_2, n_3,\ldots$ distinct sites are visited at times $t_1, t_2, t_3,\ldots$.  From this multiple-time distribution, we show that the visitation statistics of 1d random walks are temporally correlated and we quantify the non-Markovian nature of the process.  We exploit these ideas to derive unexpected results for the two-time trapping problem and also to determine the visitation statistics of two important stochastic processes, the run-and-tumble particle and the biased random walk.
\end{abstract}
\maketitle

\section{Introduction}
A key property of a diffusing particle is the territory that it covers, both because of its fundamental utility~(see, e.g., \cite{vineyard1963number,montroll1965random,Weiss:1994,Hughes:1995,feller2008introduction}), and because of its wide range of applications to diverse fields, such as chemical reactions~\cite{rosenstock1970random,donsker1979number,grassberger1982long,havlin1984trapping}, relaxation in disordered materials~\cite{klafter1985models}, and dynamics on the web~\cite{cattuto2009collective,yeung2013networking}. For a lattice random walk, the territory covered is quantified by $N(t)$, the  number of distinct sites visited by the walk up to time $t$. Underlying this  quantity is the distribution of the number of distinct sites visited at time $t$, $\mathbb{P}(N(t))$~\cite{daniels1941probability,kuhn1948aussere,feller1951asymptotic,rubin1972span,weiss1983random,redner1983asymptotic,donsker1975asymptotics,Berezhkovskii:1989}.

\begin{figure}[ht]
\includegraphics[width=0.9\columnwidth]{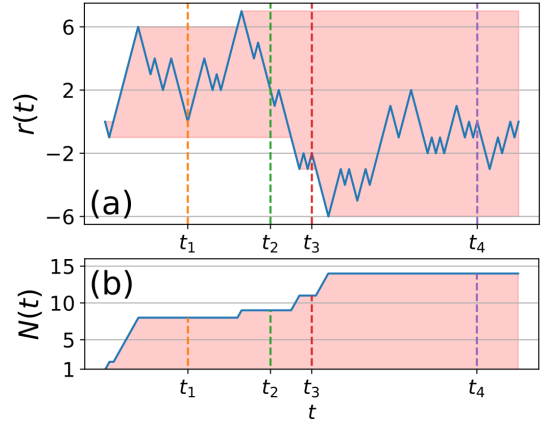}
\caption{(a) Space-time trajectory of a 1d discrete random walk  and (b) its corresponding span $N(t)$. At times $t_1,\; t_2,\; t_3$ and $t_4$ (dashed lines), we are interested in the joint statistics of $N(t_1),\; N(t_2),\; N(t_3)$ and $N(t_4)$.}
\label{definition}
\end{figure}

For the purposes of this work, it is important to emphasize that $\mathbb{P}(N(t))$ is a \emph{single-time} quantity---the distribution of $N(t)$ at \emph{one time instant}. It thus provides limited information about the full stochastic process $\{N(t)\}$, where the braces denote the set of $N(t)$ values for \emph{each} time step of the walk (Fig.~\ref{definition}).  This stochastic process is generally characterized by all its {\it multiple-time} distributions, namely the probability $\mathbb{P}(N(t_1)\!=\!n_1;\ldots;N(t_k)\!=\!n_k)$ that $n_1,n_2,\ldots,n_k$ distinct sites are visited at times $t_1<\ldots<t_k$, with $n_1\leq \ldots \leq n_k$ for any $k\geq 2$. In this Letter, we develop a methodology to determine \emph{all} these multi-time distributions analytically for 1d random walks and several fundamental generalizations.

One motivation for studying multi-time visitation distributions comes from its central role in the celebrated trapping problem~\cite{rosenstock1970random,donsker1979number,grassberger1982long,havlin1984trapping}. 
Here, a random walk wanders on a lattice that contains a fraction $c$ of immobile and randomly distributed traps, and the walk dies whenever it encounters a trap~\cite{10.1137/1.9781611971552.ch4}. The survival probability of the walk at $t$ steps, $S(t)$, equals $\langle (1-c)^{N(t)} \rangle$, where $N(t)$ is the number of distinct sites the walk visits up to $t$ steps (equivalently, the span of the walk in one dimension), and the angle brackets denote the average over all random-walk trajectories and all trap configurations. This average relies on the single-time distribution $\mathbb{P}(N(t))$.

An important extension of trapping is to the \emph{two-time} trapping problem: for a walk that has survived until time $t_1$, what is the probability $S(t_2|t_1)$ that it survives until time $t_2$? This corresponds to the probability that the walk does not encounter any traps among its newly visited sites in the time interval $[t_1,t_2]$.  Since none of the $N(t_2)-N(t_1)$ sites is a trap, this two-time survival probability is
\begin{align}
\label{Rosenstock}
S(t_2|t_1) &\equiv \left \langle (1-c)^{N(t_2)-N(t_1)} \right \rangle\\
&=\sum_{n_1 \leq n_2}(1\!-\!c)^{n_2-n_1}\mathbb{P}\left( N(t_1)\!=\!n_1;N(t_2)\!=\!n_2 \right)\,, \nonumber
\end{align}
which thus relies on the the two-time span distribution.
This trapping probability reveals a striking aging feature: if a walk survives until time $t_1$, its survival statistics at later times is strongly modified, since we now have extra information about the location of traps.  From the two-time span distribution, we will show the surprising effect that the survival probability $S(t_2|t_1)$ goes to a \emph{non-zero} value, independent of time and trap concentration, when $t_2$ is a multiple of $t_1$. 

We finally emphasize that $\{N(t)\}$ is not a Gaussian process (even the single-time distribution, $\mathbb{P}(N(t))$, is not Gaussian~\cite{daniels1941probability,kuhn1948aussere,feller1951asymptotic,rubin1972span,weiss1983random,redner1983asymptotic,Hughes:1995}), and thus is not fully characterized by the knowledge of its mean and covariance (partial results for the latter quantities are given in~\cite{PhysRevLett.117.080601,B_nichou_2016,Annesi:2019}).  Thus determining the full two-time span distribution requires new theoretical developments. We also stress that $\{N(t)\}$ is not even a Markovian process. That is, knowledge of $N(t')$ at time $t'$ is insufficient to determine the properties of $N(t)$ for $t>t'$, because the position of the random walk at time $t'$ is not known. As a consequence, not only the two-time distribution but all $k$-time distributions are needed to fully characterize the process $\{N(t)\}$.\\

\section{Single-time distribution} 

To introduce our formalism, we first show how to recover the classic asymptotic distribution of $N(t)$ for a nearest-neighbor symmetric random walk. Our approach relies on the random variable $\tau_k$, defined as the elapsed time between visits to the $k^{\rm th}$ and $(k+1)^{\rm st}$ distinct sites. Crucially, these times $\tau_0,\dots,\tau_n$ are \emph{independent} for a 1d symmetric nearest-neighbor random walk. This independence arises because the distribution of times for a random walker to visit a new site when it starts from the edge of already visited interval depends \emph{only} on the number of distinct sites already visited and nothing else.

We now relate the statistics of $N(t)$ to that of the times $\tau_k$ by noting that
\begin{eqnarray}
\label{startingpoint}
\mathbb{P} (N(t)\ge n)=\sum_{k=0}^t \mathbb{P} (\tau_0+\ldots+\tau_{n-1}=k)\,,
\end{eqnarray}
with the convention $N(0)=1$ and $\tau_0=0$. That is, to visit at least $n$ distinct sites by time $t$, the walk must visit $n$ distinct sites by time $t$ or earlier.  We now define the discrete Laplace transform for any function $f$ as
\begin{equation*}
\mathcal{L}(f(t))\equiv{\widehat  f}(s)\equiv\sum_{t=0}^{\infty} f(t) e^{-st}\,,
\end{equation*}
from which the Laplace transform of $\mathbb{P}(N(t)\ge n)$ is
\begin{align}
\label{onetime}
     \mathcal{L}\big( \mathbb{P} (N(t)\ge n) \big) 
     &=\sum_{k=0}^{\infty} \sum_{t=k}^{\infty} e^{-st} \mathbb{P} (\tau_0+\ldots+\tau_{n-1}=k) \nonumber \\
     &=\sum_{k=0}^{\infty} \frac{e^{-sk}}{1-e^{-s}}  \mathbb{P} (\tau_0+\ldots+\tau_{n-1}=k) \nonumber \\[0mm]
     &=\frac{1}{1-e^{-s}}\prod\limits_{k=0}^{n-1}\widehat{F}(s,k)\,.
\end{align}
In the first line, the sums over $k$ and $t$ have been interchanged, the last line exploits the independence of the $\tau_k$, and $\widehat{F}(s,k)$ is the Laplace transform of the exit-time distribution from an interval of length $k$, when the walk starts a unit distance from its edge~\cite{Redner:2001}. Here, exit from the interval corresponds to visiting a new site. We obtain the large-$k$ asymptotic distribution of $N(t)$ from the behavior of ${\widehat F}(s,k)$ in the limit $k \to \infty$, $s \to 0$, with $sk^2$ finite.  In Appendix~\ref{appA}, we show that 
\begin{align}
\begin{split}
\label{expansion}
{\widehat F}(s,k)&=1+g(s,k)+o(\sqrt{s}) \\
\mbox{with~ }g(s,k) &\equiv  -\sqrt{2s}\tanh \left( \sqrt{sk^2/2\,} \right)\,.
\end{split}
\end{align}
The logarithm of the product in Eq.~\eqref{onetime} is then asymptotically given by
\begin{eqnarray}
    \ln\Big[\prod_{k=0}^{n-1} {\widehat F}(s,k)\Big]
    &\sim&\int_0^n g(s,k) dk\,.
\end{eqnarray}
Substituting this result in \eqref{onetime} yields the Laplace transform of the distribution of the number of distinct sites visited:
\begin{align}
\label{Lt}
\mathcal{L}\big( \mathbb{P} (N(t)=n) \big)&=-\partial_n\mathcal{L}\left( \mathbb{P} (N(t)\ge n) \right)\nonumber\\
&\sim-\frac{1}{s}\;\partial_n \left( \frac{h(s,0)}{h(s,n)}\right)\,,
\end{align}
where 
\begin{align*}
h(s,n)\equiv \exp\left( \int_n^0 g(s,k)dk \right)=\cosh^2\left( \sqrt{sn^2/2}\right)\,.
\end{align*} 
Laplace inversion of \eqref{Lt} finally gives the well-known expression for the asymptotic distribution of the number of distinct sites visited by a 1d nearest-neighbor symmetric random walk~\cite{daniels1941probability,kuhn1948aussere,feller1951asymptotic,rubin1972span,weiss1983random,redner1983asymptotic,Hughes:1995}; this is also equivalent to the distribution of the span of a 1d Brownian motion with diffusion constant $D=1/2$ at any time.\\
   
\section{Two-time distribution} 

We now generalize and determine the multiple-time distributions of $\lbrace N(t) \rbrace$, starting with the two-time distribution. Parallel to the one-time distribution, note that for $t_1\le t_2$ and $n_1\le n_2$, we have
\begin{align}
\label{P2}
&\mathbb{P}\big(N(t_1)\ge n_1;N(t_2)\ge n_2\big) \nonumber\\[1mm]
&~=\mathbb{P}\big(\tau_0+...+\tau_{n_1-1} \leqslant t_1 ; \tau_0+...+\tau_{n_2-1} \leqslant t_2 \big) \nonumber \\
&~=\sum\limits_{k_1=0}^{t_1}\!\!\sum\limits_{k_2=0}^{t_2-k_1} \mathbb{P} (\tau_0\!+\!\ldots\!+\!\tau_{n_1-1}=k_1;\tau_{n_1}\!+\!\ldots\!+\!\tau_{n_2-1}\!=\!k_2)\,.\nonumber\\
\end{align}
Taking the (two-variable) discrete Laplace transform, exploiting the independence of the $\tau_i$, and noting that the upper bound of the second sum  depends on the argument of the first sum ($k_2 \leq t_2-k_1$), we obtain
\begin{eqnarray}
&&\mathcal{L}\big(\mathbb{P} (N(t_1)\ge n_1;N(t_2)\ge n_2) \big)=\nonumber\\ && \quad \frac{{\widehat F}(s_1\!+\!s_2,0)\ldots{\widehat F}(s_1\!+\!s_2,n_1\!-\!1){\widehat F}(s_2,n_1)\ldots{\widehat F}(s_2,n_2\!-\!1)}{(1-e^{-s_1})(1-e^{-s_2})}\, , \nonumber \\
\end{eqnarray}
where the argument $s_1+s_2$ comes from the upper bound dependency. Using  Eq.~\eqref{expansion}, we find, in the large-time (small-$s$) limit 
\begin{align}
\label{2timelaplace} 
&\mathcal{L}\big(\mathbb{P}(N(t_1) \geq n_1;N(t_2) \geq n_2) \big)\nonumber\\
&\sim\frac{1}{s_1 s_2}{\widehat F}(s_1\!+\!s_2,0)\ldots{\widehat F}(s_1\!+\!s_2,n_1\!-\!1)\nonumber\\
&\qquad\qquad\times{\widehat F}(s_2,n_1)\ldots{\widehat F}(s_2,n_2-1)  \nonumber\\
&\sim \frac{1}{s_1s_2}\;\exp{\left[\int_0^{n_1} \!\!\!g(s_1+s_2,k) dk+\int_{n_1}^{n_2}\!\!\!g(s_2,k)dk\right]} \nonumber \\
&\sim  \frac{1}{s_1s_2}\; \frac{h(s_1+s_2,0)}{h(s_1+s_2,n_1)}\frac{h(s_2,n_1)}{h(s_2,n_2)}\,,
\end{align}
for $n_1 \leq n_2$. We then Laplace invert this formula to obtain the expression for the asymptotic two-time distribution that appears in Eq.~(\ref{eq:ktimecum}) of Appendix~\ref{appA}.

\begin{figure}[!t]
\includegraphics[width=\columnwidth]{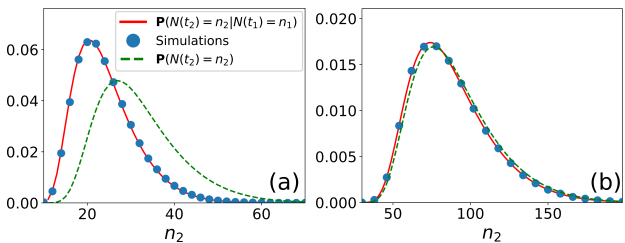}
\caption{\label{Indep} The conditional $2$-time distribution (simulations, blue dots; theory, red curve) and its convergence to the single-time distribution (dashed). Shown is $\mathbb{P}(N(t_2)\!=\!n_2|N(t_1)\!=\!n_1)$ versus $n_2$ with fixed $n_1\!=\!10$ and $t_1\!=\!200$, with (a) $t_2\!=\!400$ and (b) $t_2\!=\!3200$.} 
\end{figure}

Equation~\eqref{2timelaplace} has several important consequences: \\ (i) First, we may verify that the covariance of the span, obtained in~\cite{Annesi:2019}, follows from the complete two-time distribution Eq.~\eqref{2timelaplace} (see Appendix~\ref{appA});  (ii) Second, for $t_2/t_1 \to \infty$, with $t_i,n_i\to \infty$ and $n_i/t_i^{1/2}=a_i$ fixed for $i=1,2$, the deviation between the two-time distribution and the product of one-time distributions reduces to 
\begin{align}
&\mathbb{P}\big(N(t_1)\!=\!n_1;N(t_2)\!=\!n_2\big)\nonumber\\
&\quad -\mathbb{P}\big(N(t_1)\!=\!n_1\big)\;\mathbb{P}\big(N(t_2)\!=\!n_2\big)
 \sim C_{a_1,a_2}\;\frac{t_1^{1/2}}{t_2^{3/2}}\,, \label{difference}
\end{align}
where the expression for the constant $C_{a_1,a_2}$ is given in Eq.~\eqref{eq:Ca1a2} of Appendix~\ref{appA}. Thus temporal correlations in the two-time distribution are long range, and statistical independence of $N(t_1)$ and $N(t_2)$ is recovered only in the limit $t_1,t_2\to\infty$ with $t_2\gg t_1$; (iii) Third, we also obtain the conditional $2$-time distribution 
\begin{align*}
\mathbb{P}\big(N(t_2)\!=\!n_2\vert N(t_1)\!=\!n_1\big)=\frac{\mathbb{P}\big(N(t_2)\!=\!n_2; N(t_1)\!=\!n_1\big)}{\mathbb{P}\big( N(t_1)\!=\!n_1\big)}\,.
\end{align*}
Figure~\ref{Indep} illustrates the slow convergence of the conditional $2$-time distribution to the single-time distribution $\mathbb{P} (N(t_2)=n_2|N(t_1)=n_1) \to \mathbb{P} (N(t_2)=n_2)$ when $t_2\gg t_1$.

\section{$k$-time distributions} 

Following our theoretical approach, the Laplace transform of the $k$-time span distribution is given by (compare with Eq.~\eqref{2timelaplace})
\begin{eqnarray}
\label{laplacektime}
&& \mathcal{L}\big( \mathbb{P}\left( N(t_1) \geqslant n_1;\ldots; N(t_k) \geqslant n_k \right) \big) \nonumber \\
&\sim&\frac{1}{s_1\ldots s_k}\frac{h(s_1+\ldots+s_k,0)}{h(s_1+\ldots+s_k,n_1)}\frac{h(s_2+\ldots+s_k,n_1)}{h(s_2+\ldots+s_k,n_2)}\times\dots\nonumber\\
&&\dots\times\frac{h(s_k,n_{k-1})}{h(s_k,n_k)}\,.
\end{eqnarray}
We can derive and Laplace invert this expression to obtain the expression given by Eq.~\eqref{eq: ktimedis} of Appendix~\ref{appB}, namely, $\mathbb{P}(N(t_1) = n_1,...,N(t_k) = n_k)$. 
 
\begin{figure}[!t]
\includegraphics[width=\columnwidth]{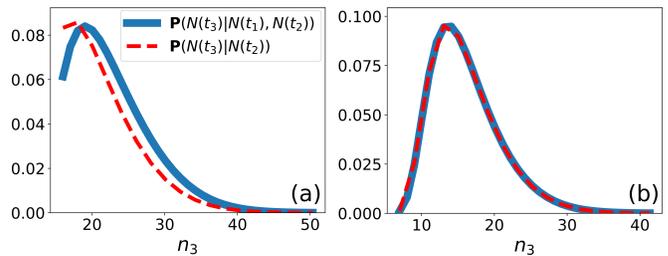}
\caption{\label{Markov} Non-Markovian property of the span $\lbrace N(t) \rbrace$. The distribution $\mathbb{P}(N(t_3)\!=\!n_3|N(t_1)\!=\!n_1;N(t_2)\!=\!n_2)$ of the quantity $N(t_3\!=\!200)$ conditioned on $N(t_1\!=\!100)\!=\!5$ and (a) $N(t_2\!=\!110)\!=\!15$ and (b) $N(t_2\!=\!110)\!=\!6$ (blue curves). The distribution $\mathbb{P}(N(t_3)\!=\!n_3|N(t_2)\!=\!n_2)$ of $N(t_3)$ conditioned only on $N(t_2)$ is represented by the red dashed curves.}
\end{figure}

We highlight the non-Markovian property of $\{N(t)\}$ by comparing $\mathbb{P}\left(N(t_3)=n_3|N(t_1)=n_1; N(t_2)=n_2 \right)$ and $\mathbb{P}\left(N(t_3)=n_3| N(t_2)=n_2 \right)$, as shown in Fig.~\ref{Markov}. For given $N(t_1)$ and $N(t_2)$, specifying both observables can change the distribution of the span at later times compared to the distribution when only $N(t_2)$ is specified. This quantifies how the distribution of visited sites at a particular time depends on previous values of $N(t)$.

We may also use Eq.~\eqref{laplacektime} to calculate the difference between the $k$-time distribution and the product of $k$ one-time distributions, analogous to Eq.~\eqref{difference}, for $t_1 \ll t_2\ll \ldots \ll t_k$
\begin{align}
\mathbb{P}&\left( N(t_1) = n_1;\ldots; N(t_k) = n_k \right) \nonumber \\
&\quad-\mathbb{P}\left(N(t_1) =  n_1 \right)\ldots \mathbb{P}\left(N(t_k) = n_k \right) \nonumber \\
& \quad\qquad\sim  \frac{1}{\sqrt{t_1\ldots t_k}} \sum\limits_{\ell=1}^{k-1}\frac{t_\ell}{t_{\ell+1}}\; C^\ell_{a_1,...,a_\ell}\,,
\end{align} 
with $a_i=n_i/\sqrt{t_i}$ fixed, and where the expression for the constant $C^\ell_{a_1,...,a_k}$ is given in Eq.~\eqref{eq:Cla1a2} of Appendix~\ref{appB}. This slow temporal decay means that span correlations between $k$ time points are long range and are controlled by the largest ratio $t_\ell/t_{\ell+1}$ between successive times. 

We emphasize that Eq.~\eqref{laplacektime} fully characterizes the stochastic process $\{N(t)\}$ in that we can compute \emph{any} functional of  $\{N(t)\}$. Two important and natural examples are general-order moments of the distribution at arbitrary time points, $\mathbb{E}\left(N(t_1)^{\alpha_1}\ldots N(t_k)^{\alpha_k} \right)$, and the joint statistics $\mathbb{P}\left(T_{n_1}=t_1; \ldots ; T_{n_k}=t_k \right)$ of the time to first visit $n$ distinct sites $T_n \equiv \min \left\lbrace t| N(t) > n  \right\rbrace$.

\section{Two-time trapping problem}
 
\begin{figure}[ht]
    \centering
    \includegraphics[width=0.8\columnwidth]{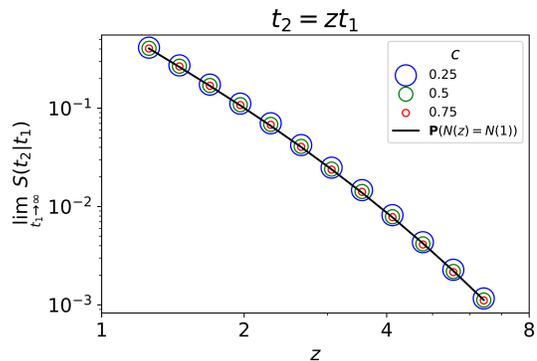}
    \caption{Long-time limit of the two-time survival probability, $S(t_2|t_1)$ (see Eq.~\eqref{Rosenstock}), of the 1d random walk at time $t_2$ knowing that it survived until time $t_1$.  Here $z=t_2/t_1$ is a constant, with $t_1,t_2\to\infty$. The black line represents the asymptotic theoretical value, Eq.~(\ref{eq:fig4}) of Appendix~\ref{appC}. Symbols correspond to numerical simulations, for different values of the fraction $c$ of immobile and randomly distributed traps.}
    \label{fig:Trap2}
\end{figure}

We can obtain the exact two-time survival probability $S(t_2|t_1)$ defined above by substituting the two-time span distribution~\eqref{2timelaplace} into Eq.~\eqref{Rosenstock}.  The explicit form for this two-time survival probability is given as Eq.~\eqref{eq:RosenstockSimpler} of Appendix~\ref{appC}. Three distinct cases arise: (a) $t_2-t_1\ll t_1$, where $S(t_2|t_1) \to 1$, (b) $t_2-t_1\gg t_1$, where $S(t_2|t_1) \to 0$, and (c) $t_1,t_2\to \infty$, with $z \equiv t_2/t_1$ a constant. While the behavior of $S(t_2|t_1)$ in the first two regimes can be qualitatively inferred from the one-time span distribution, the last case is subtle and requires the knowledge of two-time quantities.
As shown in Fig.~\ref{fig:Trap2}, the survival probability up to time $t_2$ goes to a \emph{non-zero} value that depends only on the ratio $z\equiv t_2/t_1$, and \emph{not} on the trap concentration. Since the typical size of the new territory explored beyond time $t_1$, $\left \langle N(t_2)-N(t_1) \right \rangle \propto \sqrt{Dt_1}\left(\sqrt{z} -1\right)$, diverges at large times, we would naively expect that the probability to encounter a trap among these newly visited sites will be close to 1 for $t_2\to\infty$. Thus the walk should be trapped with high probability at time $t_2$, even when conditioned to survive to time $t_1$.  

In contrast, the conditional survival probability goes to a non-zero value. This behavior corresponds to the probability for the walk to not discover any new sites in the time interval $[t_1,t_2]$, $\mathbb{P}(N(t_1)=N(t_2))$ (see Eq.~(\ref{eq:fig4}) in Appendix~\ref{appC}). Surprisingly, this expression is concentration independent for the particular choice $t_2=zt_1$, and this result reveals itself only through the two-time distribution. Thus correlations between the number of distinct sites visited at different times are crucial for understanding observables that are functionals of several of these variables.

\section{Run-and-tumble particle}

More generally, our formalism for the multiple-time span distribution can be applied to any type of random walk with: (i) a simply connected span (i.e., no ``holes" in the trajectory), (ii) translation invariance (i.e., the distribution for the next step is independent of location), (iii) symmetry, so that the exit-time statistics starting from either end of the interval of visited sites are the same. An important example is the continuous run-and-tumble particle, a classical model of bacteria motility (see, e.g., \cite{Weiss:2002,tailleur2009sedimentation,bechinger2016active,larralde2020first}). Such a particle moves ballistically at a constant speed $v$ during a flight time that is exponentially distributed with average duration $T$ (a ``run"), after which the particle ``tumbles", i.e., chooses a new direction.

For this continuous-space example, the number of distinct sites visited is replaced by the length of the span. We again define $g(s,k)$ as the leading contribution to ${\widehat F}(s,k)-1$ for $s\to 0$ (see Eq.~\eqref{expansion}). Equation~\eqref{laplacektime} still applies, with $h(s,n)$ again given by $\exp{\left(\int^n_0 g(s,k)dk \right)}$. In the SM, we show that for this process
\begin{equation}
    h(s,n)= \frac{1+\sqrt{\frac{sT}{1+sT}} \tanh \left( w n/2 \right)}{1- \sqrt{\frac{sT}{1+sT}} \tanh \left( wn/2 \right)}\frac{ sT+\cosh^2(w n/2)}{sT+1}\,, \label{eq:persist}
\end{equation}
where $w^2\equiv s^2/v^2+s/(v^2T)$. Together with Eq.~\eqref{laplacektime}, we thus obtain the Laplace transform of the $k$-time distribution for the span of a run-and-tumble particle. By numerical inversion of this Laplace transform, we can compute any multiple-time distribution, see Appendix~\ref{appF}.

For times much longer than the persistence time scale, namely, $t_i\gg T$, where $\{t_i\}$ is the set of times $t_1 \ll t_2 \ll \ldots \ll t_k$ at which the span is sampled, the run-and-tumble walk approaches a Brownian motion with diffusion constant $D=v^2T$.
The covariance of the span is shown in Appendix~\ref{appD} to have a relative correction that is proportional to ${T}/{t_1}$ compared to a pure symmetric random walk. This shows that the span of a run-and-tumble walk converges algebraically towards that of  Brownian motion. Moreover, it is the smallest of the sampling times $t_1$ that controls this relative difference.\\

\section{Biased random walk \label{biased}}

We can generalize still further to treat a biased random walk that hops one site to the right with probability $p$ or one site to the left with probability $q=1-p$ in a single step. While the previous point (iii) about the symmetry of the exit-time statistics no longer holds, it is again possible to compute the multiple-time span distributions. There are two new issues that we need to resolve to compute these distributions: (i) the asymmetry of exit-time statistics and (ii) the dependence of random variables $\tau_i$. For example, if the bias is to the right and a new site is reached at the right extremity of the visited region, then a small value of $\tau_i$ likely leads to a small value of $\tau_{i+1}$. 
These two difficulties can be overcome by introducing the coupled variables $(\tau_i, \delta_i)$,  where $\delta_i$  denotes the direction (left or right end of the visited interval) of the random walker when the site $i+1$ is first visited. We further need to replace the exit-time distributions $\widehat{F}(s,k)$ by $2\times 2$ matrices of exit-time distributions, whose first index represents the start of the walk (left or right side of the interval), while the second index represents the exit side. The Laplace transform of the $k$-time span distribution can now be expressed as
\begin{align}
\label{Lb}
&\mathcal{L} \left( \mathbb{P} (N(t_1) \geq n_1;\ldots;N(t_k) \geq n_k)\right)  \nonumber \\
&\sim \frac{1}{2s_1 s_2\ldots s_k}\;  (1,1) \times M(s_1+\ldots+s_k,0,n_1)\times\nonumber\\ 
& M(s_2+\ldots+s_k,n_1,n_2)\times \ldots\times M(s_k,n_{k-1},n_k) \times (1,1)^T \,,\nonumber \\
\end{align}
where $M(s,m,n)$ is a $2\times 2$ matrix defined in Eqs.~(\ref{eq:matrix})-(\ref{eq:nu}) of Appendix~\ref{appE}. This general $k$-time distribution reduces to the one-time distribution recently found in~\cite{wiese2020span}. After numerical Laplace inversion (see Appendix~\ref{appF}), it provides the $k$-time distributions, and thus the full characterization of the span of a biased random walk.

\section{Summary}

To summarize, we developed a new approach to compute the multiple-time distributions of the span of a one-dimensional random walk, which fully characterize the time evolution of the span of the walk.  We showed that temporal correlations in the span decay slowly, so that the span exhibits a long-time memory. We applied our formalism to uncover unexpected behavior of the two-time trapping problem and we generalized our approach to determine the multiple-time span distribution for a run-and-tumble particle and a biased random walk. 

A significant theoretical challenge is to extend our results to higher spatial dimensions. While important results are available for the single-time visitation distribution~\cite{le1986proprietes}, nothing is known for the multi-time visitation distribution. Equation~\eqref{startingpoint} holds generally and constitutes the starting point to determine multiple-time distributions of the number of distinct sites visited in any dimension $d$. However, for $d>1$, non-trivial correlations between the $\tau_i$'s arise. To deal with these correlations developing new theoretical methods is crucial.

\begin{acknowledgments}

SRs research was supported in part by NSF grant DMR-1910736.

\end{acknowledgments}


\onecolumngrid
\appendix

\renewcommand{\thefigure}{Figure S\arabic{figure}}
\renewcommand{\figurename}{}
\setcounter{figure}{0}

\section{The two-time span distribution}\label{appA}
\subsection{Derivation of Eq.~(9) in the main text}
We decompose the process $\left\lbrace N(t) \right\rbrace$ of the number of distinct visited sites at a set of times $\{t\}$ by using the fact that the times between visits to new sites, $\{ \tau_i \}$, are independent. Supposing that $n_1 \leqslant n_2$ we have
\begin{align}
\mathbb{P}\left( N(t_1) \geqslant n_1; N(t_2) \geqslant n_2 \right) &=
\sum\limits_{k=0}^{\min(t_1,t_2)}\mathbb{P}\left(\tau_0+\ldots +\tau_{n_1-1}=k; \tau_{n_1}+\ldots +\tau_{n_2-1} \leqslant t_2-k \right)  \nonumber \\[2mm]
&=\sum\limits_{k=0}^{\min(t_1,t_2)}\mathbb{P}\left(\tau_0+\ldots +\tau_{n_1-1}=k \right) \mathbb{P}\left( \tau_{n_1}+\ldots +\tau_{n_2-1} \leqslant t_2-k \right).
\end{align}
Performing the discrete Laplace transform, 
$\displaystyle \mathcal{L}(f(t)) = \sum_{t\geq 0} f(t)e^{-st} \equiv \widehat{f}(s)$,
on both the variables $t_1$ and $t_2$ gives
\begin{align} 
\label{eq:MainCalcul}
\mathcal{L}&\left( \mathbb{P}\left( N(t_1) \geqslant n_1; N(t_2) \geqslant n_2 \right) \right) 
=\sum\limits_{t_1,t_2=0}^{\infty}\sum\limits_{k=0}^{\min(t_1,t_2)}e^{-s_1 t_1}\;e^{-s_2 t_2}\;\mathbb{P}\left(\tau_0+\ldots +\tau_{n_1-1}=k \right) \mathbb{P}\left( \tau_{n_1}+\ldots +\tau_{n_2-1} \leqslant t_2-k \right)  \nonumber\\
&\qquad =
\frac{1}{1-e^{-s_1}}\sum\limits_{0 \leq k \leq t_2 \leq \infty}e^{-(s_1+s_2)k}\;\mathbb{P}\left(\tau_0+\ldots +\tau_{n_1-1}=k \right)\; e^{-s_2(t_2-k)}\;\mathbb{P}\left( \tau_{n_1}+\ldots +\tau_{n_2-1} \leqslant t_2-k \right) \nonumber \\
&\qquad =\frac{1}{(1-e^{-s_1})(1-e^{-s_2})}\;\widehat{F}(s_1+s_2,0)\ldots \widehat{F}(s_1+s_2,n_1-1)\widehat{F}(s_2,n_1)\ldots \widehat{F}(s_2,n_2-1) \, , 
\end{align}
where $\widehat F(s,k)$ is the Laplace transform of the exit-time probability from an interval of length $k$ when a diffusing particle starts a distance $1$ from the edge of the interval, see Eq.~(2.2.10) of~\cite{Redner:2001} (noting that $\widehat F(s,0)=1$ as $\tau_0=0$). This probability is
\begin{align}
\widehat F(s,k) &=\frac{\sinh\left(\sqrt{2s} \right)+\sinh\left(\sqrt{2s}(k-1) \right)}{\sinh\left(\sqrt{2s}k \right)}\nonumber\\
&=1-\sqrt{2s}\tanh \left( \sqrt{sk^2/2\,} \right)+o\left( \sqrt{s}\right)\equiv 1+ g(s,k)+o\left( \sqrt{s}\right) \, ,
\end{align}
where the limit $s \to 0$ and $\sqrt{sk^2}$ fixed is taken in the second line.
For $s_1,s_2 \to 0$, we obtain the Laplace transform of the distribution, $\mathcal{L}\left( \mathbb{P}\left( N(t_1) \geqslant n_1; N(t_2) \geqslant n_2 \right) \right)$, as
\begin{align}
\mathcal{L}\left( \mathbb{P}\left( N(t_1) \geqslant n_1; N(t_2) \geqslant n_2 \right) \right) &\sim \frac{1}{s_1s_2}\exp \left( \int_0^{n_1}g(s_1+s_2,k)dk+\int_{n_1}^{n_2}g(s_2,k)dk\right) \nonumber\\[2mm]
&= \frac{1}{s_1s_2}\;\frac{1}{\cosh^2\left(n_1\sqrt{(s_1+s_2)/2}\right)}\;\frac{\cosh^2\left(n_1\sqrt{s_2/2}\right)}{\cosh^2\left(n_2\sqrt{s_2/2}\right)}. \label{eq:twotimes}
\end{align}

In the following, subscripts of the Laplace transform $\mathcal{L}$ are precised only when the variable in subscript is the only one which has been Laplace transformed. Otherwise, all the time variables are Laplace transformed. 

We now want the inverse Laplace transform of the quantity \eqref{eq:twotimes}. First, we perform the inverse transform on the variable $s_1$. We use the residue theorem to compute the complex integral coming from the inverse Laplace transform. 
\begin{align}
&\mathcal{L}_{t_2 \to s_2}\left( \mathbb{P}\left( N(t_1) \geqslant n_1, N(t_2) \geqslant n_2 \right) \right) \nonumber \\
& \quad= \int_{\gamma-i \infty}^{\gamma+i\infty} \frac{dz}{2 \pi i} e^{zt_1}\frac{1}{zs_2}\;\frac{1}{\cosh^2\left(n_1\sqrt{(z+s_2)/2}\right)}\;\frac{\cosh^2\left(n_1\sqrt{s_2/2}\right)}{\cosh^2\left(n_2\sqrt{s_2/2}\right)} \\
& \quad= \frac{\cosh^2\left(n_1\sqrt{s_2/2}\right)}{s_2\cosh^2\left(n_2\sqrt{s_2/2}\right)} \frac{\partial}{\partial n_1} \left[ \int_{\gamma-i \infty}^{\gamma+i\infty} \frac{dz}{2 \pi i} \frac{e^{zt_1}}{z}\;\sqrt{\frac{2}{(z+s_2)}}\tanh\left(n_1\sqrt{\frac{z+s_2}{2}}\right) \; \right] \\
&\quad= \frac{1}{s_2}\frac{\cosh^2\left(n_1\sqrt{\frac{s_2}{2}}\right)}{\cosh^2\left(n_2\sqrt{\frac{s_2}{2}}\right)}\frac{\partial}{\partial n_1}   \sum_ {k=0}^{\infty}\text{Res}\left\{\frac{e^{zt_1}}{z}\sqrt{\frac{2}{(z+s_2)}}\tanh \left( n_1 \sqrt{\frac{z+s_2}{2}} \right) ,
~z=-\frac{2}{n_1^2}\left( \frac{\pi}{2}+k\pi\right)^2-s_2\right\}  \nonumber \\
& \qquad\qquad +\frac{1}{s_2\cosh^2\left(n_2\sqrt{\frac{s_2}{2}}\right)}  \\
&\quad = -\frac{1}{s_2}\frac{\cosh^2\left(n_1\sqrt{s_2/2}\right)}{\cosh^2\left(n_2\sqrt{s_2/2}\right)}\frac{\partial}{\partial n_1} \left\{\sum_ {k=0}^{\infty}e^{-\frac{2 \pi^2}{n_1^2}(k+1/2)^2t_1-s_2t_1}\frac{4}{\left(s_2+\frac{2}{n_1^2}\left( \frac{\pi}{2}+k\pi\right)^2\right)n_1}\right\} +  \frac{1}{s_2\cosh^2\left(n_2\sqrt{s_2/2}\right)} \; ,
\end{align}
$\gamma$ being a positive number such that the real part of any poles is smaller than $\gamma$. In the rest of the text, we will take $\gamma \to 0^+$ as all poles have real parts $\leq 0$.
Performing the inverse Laplace transform $s_2 \to t_2$ in a similar manner, we obtain 
\begin{align}
&\mathbb{P}\left( N(t_1) \geqslant n_1; N(t_2) \geqslant n_2 \right) \nonumber \\
&\quad =\int_{-i \infty}^{+i\infty} \frac{dz}{2 \pi i} e^{zt_2} \frac{\sqrt{2}}{z^{3/2}}\partial_{n_2}\Bigg\{ -\cosh^2\left(n_1\sqrt{\frac{z}{2}}\right)\tanh\left(n_2\sqrt{\frac{z}{2}}\right)\frac{\partial}{\partial n_1} \left[\sum_ {k=0}^{\infty}e^{-\frac{2 \pi^2}{n_1^2}(k+1/2)^2t_1-zt_1}\frac{4}{\left(z+\frac{2}{n_1^2}\left( \frac{\pi}{2}+k\pi\right)^2\right)n_1}\right] \nonumber \\
& \qquad\qquad + \tanh\left(n_2\sqrt{\frac{z}{2}}\right) \Bigg\}\, \\
&\quad = \frac{\partial}{\partial n_2}
\Bigg\{ \left( \frac{2}{\pi ^2} \right)^2 \sum\limits_{k'=0}^{\infty} \frac{n_2}{(k'+1/2)^2}e^{-\frac{2\pi^2}{n_2^2}(k'+1/2)^2(t_2-t_1)} \cos^2\left( \frac{n_1}{n_2}\left(\pi/2+k'\pi \right)\right) \times \nonumber \\
& \qquad \frac{\partial}{\partial n_1}\left[
\sum\limits_{k=0}^{\infty} \frac{n_1}{(k+1/2)^2-n_1^2/n_2^2(k'+1/2)^2}e^{-\frac{2\pi^2}{n_1^2}(k+1/2)^2t_1}
\right]
\Bigg\}+  \mathbb{P}\left( N(t_1) \geqslant n_1 \right) +\mathbb{P}\left(  N(t_2) \geqslant n_2 \right) -1\;, \label{eq:ktimecum}
\end{align}
with $\mathbb{P}\left( N(t) \geqslant n \right)$ which can be found in \cite{wiese2020span}. Thus we find for $n_1<n_2$:
\begin{align}
&\mathbb{P}\left( N(t_1) = n_1; N(t_2) = n_2 \right) 
= \frac{\partial^3}{\partial n_2^2 \partial n_1}
\Bigg\{ \left( \frac{2}{\pi ^2} \right)^2 \sum\limits_{k'=0}^{\infty} \frac{n_2}{(k'+1/2)^2}e^{-\frac{2\pi^2}{n_2^2}(k'+1/2)^2(t_2-t_1)} \cos^2\left( \frac{n_1}{n_2}\left(\pi/2+k'\pi \right)\right) \times
\nonumber \\
& \times \frac{\partial}{\partial n_1}\left[
\sum\limits_{k=0}^{\infty} \frac{n_1}{(k+1/2)^2-n_1^2/n_2^2(k'+1/2)^2}e^{-\frac{2\pi^2}{n_1^2}(k+1/2)^2t_1}
\right]
\Bigg\} \; . \label{eq:twotime}
\end{align}
The case $n_1=n_2$ can be obtained from Eqs.~(\ref{eq:ktimecum}) and \eqref{eq:twotime}, 
\begin{align}
\mathbb{P}&\left( N(t_1)=n_1;N(t_2) =n_1\right) =\mathbb{P}\left( N(t_1)=n_1\right)-\mathbb{P}\left( N(t_1)=n_1,N(t_2) >n_1\right) \nonumber\\
&\quad= \lim\limits_{n_2 \to n_1^+}\; \frac{\partial^2}{\partial n_2 \partial n_1}
\Bigg\{ \left( \frac{2}{\pi ^2} \right)^2 \sum\limits_{k'=0}^{\infty} \frac{n_2}{(k'+1/2)^2}e^{-\frac{2\pi^2}{n_2^2}(k'+1/2)^2(t_2-t_1)} \times  \nonumber \\
& \qquad \times \cos^2\left( \frac{n_1}{n_2}\left(\pi/2+k'\pi \right)\right)
\frac{\partial}{\partial n_1}  \left[
\sum\limits_{k=0}^{\infty} \frac{n_1}{(k+1/2)^2-n_1^2/n_2^2(k'+1/2)^2}e^{-\frac{2\pi^2}{n_1^2}(k+1/2)^2t_1}
\right]
\Bigg\} \; , \label{eq:equality}
\end{align}
whose Laplace transform in both time variables is
\begin{align}
\mathcal{L}&\left(\mathbb{P}\left( N(t_1)=n_1;N(t_2) =n_1\right)\right) \nonumber \\
&=\qquad\frac{\sqrt{2s_2}\tanh \left( n_1\sqrt{\frac{s_2}{2}}\right)+\sqrt{2s_1}\tanh \left( n_1\sqrt{\frac{s_1}{2}}\right)-\sqrt{2(s_1+s_2)}\tanh \left( n_1\sqrt{\frac{s_1+s_2}{2}}\right)}{s_1s_2\cosh^2\left(n_1\sqrt{\frac{s_1+s_2}{2}} \right)} \; \label{eq:equalLap}.
\end{align}

\subsection{Covariance of the span \label{secGleb}}
Here we show that the Laplace transform of the covariance of the span of a 1d Brownian motion obtained in \cite{Annesi:2019} is retrieved in our formalism. We note the span process of the 1d Brownian motion ie the length of the visited domain at time $t$ as $ N_D(t)$. We compare our result directly with the formula for the covariance as a function of time obtained in \cite{Annesi:2019} (taking  the diffusion coefficient $D=1/2$),
\begin{align}
\label{H12}
H(s_1,s_2) &\equiv \mathbb{E}\left( \widehat{N}_D(s_1)\widehat{N}_D(s_2) \right) \nonumber\\
&= \iint dt_1 dt_2 \;e^{-s_1t_1 -s_2t_2} \max \left( t_1,t_2 \right) \left\{ \frac{6}{\pi}\sqrt{z(1-z)}-1+\frac{2}{\pi}\arcsin\left( \sqrt{z}\right) 
+\frac{2}{\pi}\frac{z^{3/2}}{\sqrt{1-z}}g_1\left( \sqrt{\frac{z}{1-z}}\right)  \right. \nonumber \\
& \qquad
\left. \qquad+\frac{1}{\pi}\left[  \left(i+\sqrt{\frac{z}{1-z}} \right) g_1\left( i+\sqrt{\frac{z}{1-z}} \right) + \left( -i+\sqrt{\frac{z}{1-z}} \right) g_1\left( -i+\sqrt{\frac{z}{1-z}} \right) \right] \right\}\,,
\end{align}
with
\begin{align*}
z \equiv \min \left( \frac{t_1}{t_2}, \frac{t_2}{t_1} \right)\qquad \text{and} \qquad g_1 \left( \phi \right) \equiv \frac{1}{\phi^2}\int_0^1 \frac{d\beta}{\beta^2}\left( 1-\frac{\pi \phi \beta}{\sinh \left( \pi \phi \beta \right)}\right)\,.
\end{align*}

\begin{figure}[ht!]
	\includegraphics[scale=0.45]{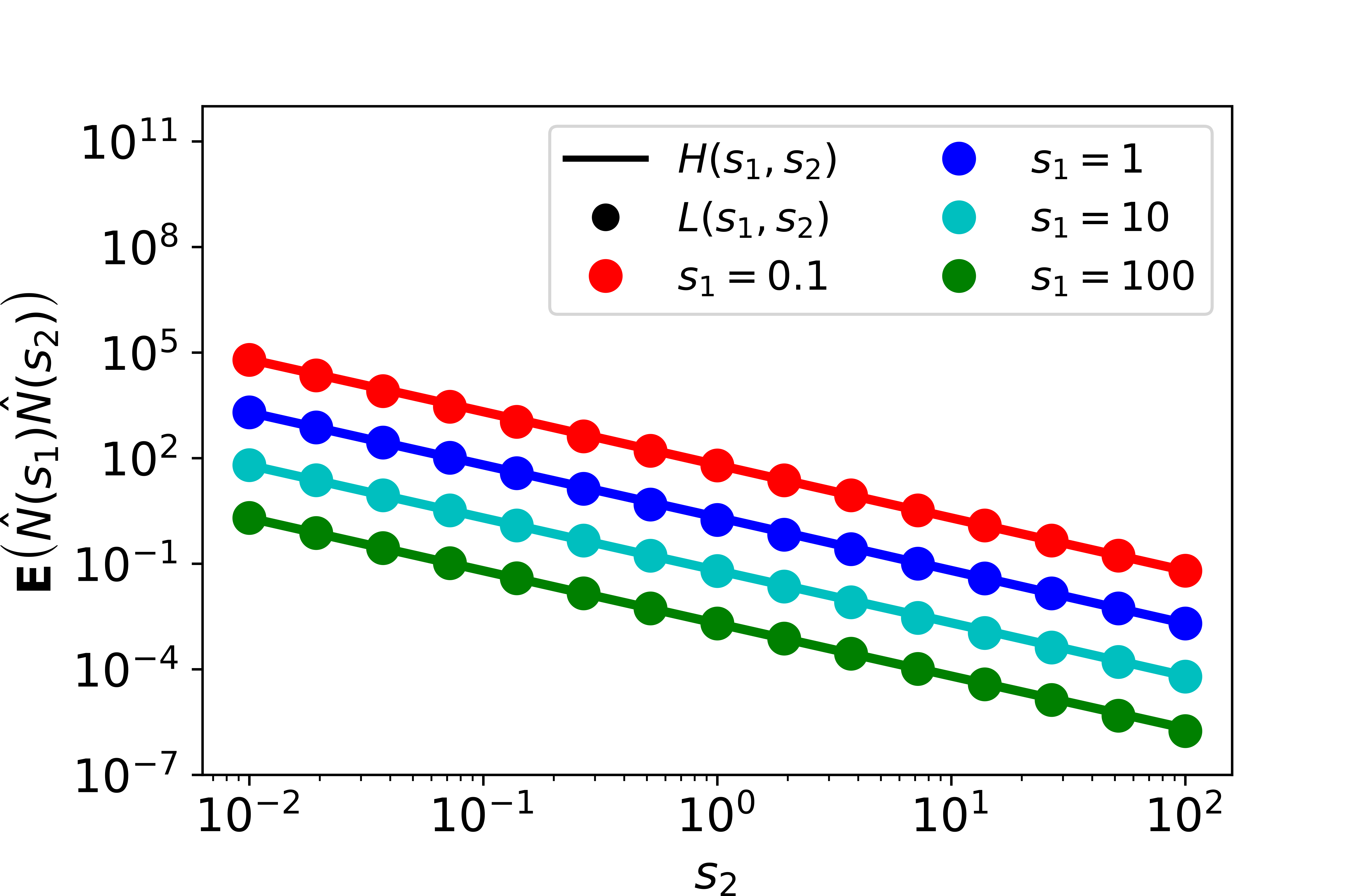}
	\caption{\label{Gleb}Comparison of the Laplace transform of the two point expectation, $H(s_1,s_2)$, obtained in \cite{Annesi:2019} and the same quantity $L(s_1,s_2)$ obtained using our formalism, Eq.~(\ref{eq:L}). }
\end{figure}

We compare the result in Eq.~\eqref{H12} from~\cite{Annesi:2019} with the one obtained from our Eqs.~(\ref{eq:twotimes}) and \eqref{eq:equalLap} by splitting the integral into the domains $n_1<n_2$, $n_1=n_2$, and $n_2<n_1$:
\begin{align}
L(s_1,s_2) &\equiv \ell(s_1,s_2)+m(s_1,s_2)+\ell(s_2,s_1), \label{eq:L}  
\end{align}
with
\begin{align}
\ell(s_1,s_2)&=\frac{1}{s_1s_2} \int_{n_1 < n_2} dn_1\, dn_2\, n_1 \,n_2 \;\frac{\partial ^2}{\partial n_1 \partial n_2} \left[ \frac{1}{\cosh^2
	\left(n_1\sqrt{\frac{s_1+s_2}{2}}\right)}\frac{\cosh^2\left(n_1\sqrt{\frac{s_2}{2}}\right)}{\cosh^2\left(n_2\sqrt{\frac{s_2}{2}}\right)} \right] \nonumber \\ 
&= \frac{2}{s_1s_2^2}\int_{0}^{ \infty} du \left(-u+u\tanh(u)-\frac{u^2}{\cosh^2(u)} \right) \frac{\partial}{\partial u } \left[ \frac{\cosh^2\left(u\right)}{\cosh^2
	\left(u\sqrt{\frac{s_1}{s_2}+1}\right)} \right]
\end{align}
and 
\begin{align}
m(s_1,s_2)= \int_{0}^{ \infty}  n^2 \;\frac{\sqrt{2s_2}\tanh \left( n\sqrt{\frac{s_2}{2}}\right)+\sqrt{2s_1}\tanh \left( n\sqrt{\frac{s_1}{2}}\right)-\sqrt{2(s_1+s_2)}\tanh \left( n\sqrt{\frac{s_1+s_2}{2}}\right)}{s_1s_2\cosh^2\left(n\sqrt{\frac{s_1+s_2}{2}}\right)}\,dn \; .
\end{align}
The numerically calculated functions $H$ and $L$ coincide as shown in \ref{Gleb}.

\subsection{Derivation of Eq.~(10) in the main text}
We compute the difference between the distributions $\mathbb{P}(N(t_1)=n_1;N(t_2)=n_2)$ and $\mathbb{P}(N(t_1)=n_1)\times\mathbb{P}(N(t_2)=n_2)$ in  the limit $t_i,n_i\to \infty$ and $n_i/t_i^{1/2}=a_i$ fixed for $i=1$ and $2$, and $n_1/n_2 \to 0$. Starting from Eq.~(\ref{eq:twotime}), we have
\begin{align}
&\mathbb{P}(N(t_1)=n_1;N(t_2)=n_2) \nonumber \\
&\quad\sim \frac{\partial^3}{\partial n_2^2\partial n_1}
\Bigg\{ \left( \frac{2}{\pi ^2} \right)^2 \sum\limits_{k'=0}^{\infty} \frac{n_2}{(k'+1/2)^2}e^{-\frac{2\pi^2}{n_2^2}(k'+1/2)^2t_2}\left[1+t_1\frac{2\pi^2}{n_2^2}(k'+1/2)^2 \right] \left[ 1-\left(\frac{n_1}{n_2}\pi\left(k'+1/2 \right)\right)^2\right]
\times \nonumber \\
& \qquad\qquad\times \frac{\partial}{\partial n_1}\left[
\sum\limits_{k=0}^{\infty} \frac{n_1}{(k+1/2)^2}\left(1+\left(\frac{(k'+1/2)n_1}{(k+1/2)n_2}\right)^2 \right)e^{-\frac{2\pi^2}{n_1^2}(k+1/2)^2t_1}
\right]
\Bigg\} \nonumber \\
&\quad\sim  \mathbb{P}(N(t_1)=n_1)\times \mathbb{P}(N(t_2)=n_2)+C_{a_1,a_2}\frac{t_1^{1/2}}{t_2^{3/2}} \label{eq:firstorder}
\end{align}
with 
\begin{align}
C_{a_1,a_2} &\equiv \frac{\partial^3}{\partial a_2^2\partial a_1}
\Bigg\{\left( \frac{2}{\pi ^2} \right)^2 \sum\limits_{k'=0}^{\infty} \pi^2 \frac{2-a_1^2}{a_2}  e^{-\frac{2\pi^2}{a_2^2}(k'+1/2)^2} 
\frac{\partial}{\partial a_1} \left[
\sum\limits_{k=0}^{\infty} \frac{a_1}{(k+1/2)^2}e^{-\frac{2\pi^2}{a_1^2}(k+1/2)^2}
\right]
\Bigg\} \nonumber \\
&\qquad+\frac{\partial^3}{\partial a_2^2\partial a_1}
\Bigg\{ \left( \frac{2}{\pi ^2} \right)^2 \sum\limits_{k'=0}^{\infty} e^{-\frac{2\pi^2}{a_2^2}(k'+1/2)^2} 
\frac{\partial}{\partial a_1}  \left[
\sum\limits_{k=0}^{\infty} \frac{a_1^3}{(k+1/2)^4 a_2}e^{-\frac{2\pi^2}{a_1^2}(k+1/2)^2}
\right]
\Bigg\}\,.\label{eq:Ca1a2}
\end{align}

\section{$k$-time span distributions}\label{appB}
\subsection{Derivation of Eq.~(11) in the main text}

For the probability distribution of the events $\lbrace N(t_1) \geqslant n_1, \ldots, N(t_k) \geqslant n_k\rbrace$, its multivariate Laplace transform is obtained by performing the same steps as those that led to (\ref{eq:MainCalcul}),
\begin{align}
& \mathcal{L}\left( \mathbb{P}\left( N(t_1) \geqslant n_1;\ldots ; N(t_k) \geqslant n_k \right) \right) \nonumber \\
&\qquad=\frac{1}{s_1\ldots s_k}\,\widehat{F}(s_1+\ldots +s_k,0)\ldots\widehat{F}(s_1+\ldots +s_k,n_1-1)\widehat{F}(s_2+\ldots +s_{k},n_1)\ldots\widehat{F}(s_k,n_k-1) \nonumber \\
&\qquad=\frac{1}{s_1\ldots s_k}\frac{1}{\cosh^2\left(n_1\sqrt{\frac{s_1+\ldots +s_k}{2}}\right)}\frac{\cosh^2\left(n_1\sqrt{\frac{s_2+\ldots +s_k}{2}}\right)}{\cosh^2\left(n_2\sqrt{\frac{s_2+\ldots +s_k}{2}}\right)}\ldots \frac{\cosh^2\left(n_{k-1}\sqrt{\frac{s_k}{2}}\right)}{\cosh^2\left(n_{k}\sqrt{\frac{s_k}{2}}\right)}. \label{ktimedis}
\end{align}

The formula for the $k$-time span distribution in the time domain is given similarly to (\ref{eq:twotime}) by  recursively performing the inverse Laplace transforms, first on $s_1$, then on $s_2$, \ldots until $s_k$, (here supposing that $n_1 < n_2 < \ldots  < n_k$),
\begin{eqnarray}
\label{eq: ktimedis}
&&\mathbb{P}\left( N(t_1) = n_1, \ldots ,N(t_k) = n_k \right) \nonumber \\
&=&\partial_{n_1,\ldots ,n_k} \left( \frac{-2}{\pi ^2} \right)^k \frac{\partial}{\partial n_k}
\Bigg[ \sum\limits_{i_k=0}^{\infty} \frac{n_k}{(i_k+1/2)^2}e^{-\frac{2\pi^2}{i_k^2}(i_k+1/2)^2(t_k-t_{k-1})} \cos^2\left( \frac{n_{k-1}}{n_k}\left(\pi/2+i_k\pi \right)\right) \times
\nonumber \\
& & \times  \frac{\partial}{\partial n_{k-1}}\Bigg[\left(
\sum\limits_{i_{k-1}=0}^{\infty} \frac{n_{k-1} \cos^2\left( \frac{n_{k-2}}{n_{k-1}}\left(\pi/2+i_{k-1}\pi \right)\right)}{(i_{k-1}+1/2)^2-n_{k-1}^2/n_k^2(i_k+1/2)^2}e^{-\frac{2\pi^2}{n_{k-1}^2}(i_{k-1}+1/2)^2(t_{k-1}-t_{k-2})}
\right) \times \nonumber \\
& &\times  \ldots  \frac{\partial }{\partial n_1} \left( \sum\limits_{i_1=0}^{\infty} \frac{n_1}{(i_1+1/2)^2-n_1^2/n_2^2(i_2+1/2)^2}e^{-\frac{2\pi^2}{n_1^2}(i_1+1/2)^2t_1}
\right)\Bigg]\ldots \Bigg] \label{eq:ktimedis}.
\end{eqnarray} 
Similarly to Eq.~\eqref{eq:equality}, the case $n_i=n_{i+1}$ for any $i$ follows from Eq.~\eqref{eq:ktimedis}.  
Additionally, we obtain the expression for the $k$-time span distribution for a Brownian motion with arbitrary diffusion constant $D$ by noting that the previous result pertains to the particular choice $D=1/2$.

\subsection{Derivation of Eq.~(12) in the main text}
We treat the limit $1 \ll t_1 \ll t_2 \ll \ldots  \ll t_k$; i.e., $s_1 \gg s_2 \gg \ldots  \gg s_k \gg 1$. Starting from Eq.~(\ref{ktimedis}), with $n_i\sqrt{s_i}=a_i$ fixed, we have 
\begin{align}
&\mathcal{L}\left( \mathbb{P}\left( N(t_1) \geq n_1,\ldots , N(t_k) \geq n_k \right) \right) \nonumber \\
&\quad =\frac{1}{s_1\ldots s_k} \frac{1}{\cosh^2\left(a_1\sqrt{\frac{s_1+\ldots +s_k}{2s_1}}\right)}\frac{\cosh^2\left(a_1\sqrt{\frac{s_2+\ldots +s_k}{2s_1}}\right)}{\cosh^2\left(a_2\sqrt{\frac{s_2+\ldots +s_k}{2s_2}}\right)}\ldots \frac{\cosh^2\left(a_{k-1}\sqrt{\frac{s_k}{2s_{k-1}}}\right)}{\cosh^2\left(\frac{a_k}{\sqrt{2}}\right)}  \nonumber \\
&\quad\sim\frac{1}{s_1\ldots s_k} \frac{1}{\cosh^2\left(a_1\sqrt{\frac{s_1+s_2}{2s_1}}\right)}\frac{\cosh^2\left(a_1\sqrt{\frac{s_2}{2s_1}}\right)}{\cosh^2\left(a_2\sqrt{\frac{s_2+s_3}{2s_2}}\right)}\ldots \frac{\cosh^2\left(a_{k-1}\sqrt{\frac{s_k}{2s_{k-1}}}\right)}{\cosh^2\left(\frac{a_k}{\sqrt{2}}\right)}  \nonumber\\[2mm]
&\quad\sim\mathcal{L}\big( \mathbb{P}\left( N(t_1) \geq  n_1 \right)\ldots  \mathbb{P}\left(N(t_k) \geq n_k \right) \big)\left[1+\sum\limits_{\ell=1}^{k-1}\frac{s_{\ell+1}}{s_{\ell}}\left(\frac{a_\ell^2}{2}-\tanh \left(\frac{a_\ell}{\sqrt{2}}  \right)\frac{a_\ell}{\sqrt{2}} \right)+o(s_{\ell+1}/s_\ell) \right]\,. 
\end{align}
Thus we obtain a result similar to the $2$-time span distribution for the first-order correction to the product of independent one-time span distributions:
\begin{align}
&\mathcal{L}\left( \mathbb{P}\left( N(t_1) = n_1;\ldots ; N(t_k) = n_k \right)-\mathbb{P}\left( N(t_1) =  n_1 \right)\ldots  \mathbb{P}\left(N(t_k) = n_k \right) \right) \nonumber \\
&\qquad\sim \frac{1}{\sqrt{s_1\ldots s_k}}  \partial_{a_1,\ldots ,a_k}\;\left[\sum\limits_{\ell=1}^{k-1}\frac{s_{\ell+1}}{s_{\ell}}\left(\frac{a_\ell^2}{2}-\tanh \left(\frac{a_\ell}{\sqrt{2}}  \right)\frac{a_\ell}{\sqrt{2}} \right)\frac{1}{\cosh^2(a_1/\sqrt{2})\ldots \cosh^2(a_k/\sqrt{2})} \right]\,.
\end{align}
Consequently, with $a_i=n_i/\sqrt{t_i}$, and using the Tauberian theorem,
\begin{eqnarray}
\mathbb{P}\left( N(t_1) = n_1;\ldots ; N(t_k) = n_k \right)-\mathbb{P}\left( N(t_1) =  n_1 \right)\ldots  \mathbb{P}\left(N(t_k) = n_k \right) 
\sim \frac{1}{\sqrt{t_1\ldots t_k}} \sum\limits_{\ell=1}^{k-1}\frac{t_\ell}{t_{\ell+1}} C^\ell_{a_1,\ldots ,a_\ell}.
\end{eqnarray} 
Here $C^\ell_{a_1,\ldots ,a_\ell}$ can be obtained either by Laplace inversion of the $\ell^{th}$ term of the sum, or starting directly from (\ref{eq: ktimedis}) and keeping the first order term in $t_\ell/t_{\ell+1} \ll 1$,
\begin{align}
& C^\ell_{a_1,\ldots ,a_k} =\mathbb{P}(N_D(1)=a_1)\mathbb{P}(N_D(1)=a_2) \ldots  \mathbb{P}(N_D(1)=a_{\ell-1}) \left( \frac{2}{\pi ^2} \right)^2   \nonumber \\
& \quad  \times \partial_{a_\ell,a_{\ell+1},a_{\ell+1}} \Bigg\{ \sum\limits_{i_{\ell+1}=0}^{\infty}  \frac{2-a_{\ell}^2}{a_{\ell+1}}\pi^2 e^{-\frac{2\pi^2}{a_{\ell+1}^2}(i_{\ell+1}+1/2)^2}
\frac{\partial}{\partial a_\ell} \left[
\sum\limits_{i_\ell=0}^{\infty} \frac{a_\ell}{(i_\ell+1/2)^2}e^{-\frac{2\pi^2}{a_\ell^2}(i_\ell+1/2)^2}
\right] \nonumber \\
&\quad  +e^{-\frac{2\pi^2}{a_{\ell+1}^2}(i_{\ell+1}+1/2)^2} 
\frac{\partial}{\partial a_\ell} \left[
\sum\limits_{i_\ell=0}^{\infty} \frac{a_\ell^3}{(i_\ell+1/2)^4 a_{\ell+1}}e^{-\frac{2\pi^2}{a_\ell^2}(i_\ell+1/2)^2}
\right] \Bigg\}
\mathbb{P}(N_D(1)=a_{\ell+2})\ldots \mathbb{P}(N_D(1)=a_{k}) \; ,\label{eq:Cla1a2} 
\end{align} 
$N_D(t)$ being the span process of the Brownian motion of parameter $D$ as defined in Appendix \ref{secGleb}.

\section{Two-time trapping problem}\label{appC}

Starting from Eq.~(1) of the main text, and going to the continuum limit, we have
\begin{align}
S(t_2|t_1)&=\left\langle(1-c)^{N(t_2)-N(t_1)}\right\rangle=\int_{n_1 \leq n_2}dn_1 dn_2\; (1-c)^{n_2-n_1}\;\mathbb{P}\left( N(t_1)=n_1,N(t_2)=n_2 \right) \\
&= \mathbb{P}(N(t_1)=N(t_2))+\int_{n_1 < n_2}dn_1 dn_2\; (1-c)^{n_2-n_1}\; \frac{\partial^2}{\partial n_2 \partial n_1}\mathbb{P}(N(t_1)\geq n_1, N(t_2) \geq n_2) \; ,\label{eq:RosenstockSimpler}
\end{align}
We now discuss the different regimes:  
\begin{enumerate}[label=(\alph*)]
	\item $t_2-t_1 \gg t_1$: In this limit, $\mathbb{P}(N(t_1)=n_1;N(t_2)=n_2) \sim \mathbb{P}(N(t_1)=n_1)\times \mathbb{P}(N(t_2)=n_2)$ for $n_1<n_2$, and $\mathbb{P}(N(t_1)=N(t_2))$ is a decreasing exponential at large times (since this quantity is related to the exit probability from an interval). Thus,
	\begin{equation}
	S(t_2|t_1) \sim \int_{n_1<n_2} dn_1 dn_2 (1-c)^{n_2-n_1} \mathbb{P}(N(t_1)=n_1)\times\mathbb{P}(N(t_2)=n_2)
	\sim f(t_1) S(t_2),
	\end{equation}
	where $f(t_1)$ depends only on $c$ and $t_1$ but not on $t_2$.
	\item $t_2-t_1 \ll t_1$:  From the scale invariance of Brownian motion, as the law of $\left(N_D(t_1),N_D(t_2) \right)$ is the same as the one of $\left(N_D(1)\sqrt{t_1},N_D(t_2/t_1)\sqrt{t_1} \right)$, the term $\mathbb{P}(N(t_1)=N(t_2)) \sim \mathbb{P}(N_D(1)\sqrt{t_1}=N_D(t_2/t_1)\sqrt{t_1})= \mathbb{P}\left(N_D(1)=N_D\left(1+\frac{t_2-t_1}{t_1}\right)\right) \to 1$ dominates. We note that $1-\mathbb{P}\left(N_D(1)=N_D\left(1+\frac{t_2-t_1}{t_1}\right)\right)$ corresponds to the probability of exiting an interval of unit size after a time $t=\frac{t_2-t_1}{t_1} \ll 1$, whose scaling behaviour is known to be $\propto \sqrt{t}$, Eq.~(5.219) of \cite{Weiss:1994}. Thus,
	\begin{eqnarray}
	1-S(t_2|t_1) &\sim& 1-\mathbb{P}\big(N_D(t_1)=N_D(t_2)\big) \propto \sqrt{\frac{t_2-t_1}{t_1}}\,.
	\end{eqnarray}
	\item $t_2=zt_1$: The term which dominates is still $\mathbb{P}(N(t_1)=N(t_2)) \sim \mathbb{P}(N_D(1)=N_D\left( z\right))$, as the integral in $n_1<n_2$ is decreasing (at least as a stretched exponential). Moreover, the conditional survival probability $S(t_2|t_1)$ converges to a value that is neither 0 nor 1: 
	\begin{align}
	S(t_2|t_1) &\to \mathbb{P}(N_D(1)=N_D\left( z\right)) \nonumber \\
	&\qquad=\int_0^{\infty} dn_1 \lim\limits_{n_2 \to n_1^+} \frac{\partial^2}{\partial n_2 \partial n_1}
	\Bigg\{ \left( \frac{2}{\pi ^2} \right)^2 \sum\limits_{k'=0}^{\infty} \frac{n_2\cos^2\left( \frac{n_1}{n_2}\left(\pi/2+k'\pi \right)\right)}{(k'+1/2)^2}\;e^{-\frac{2\pi^2}{n_2^2}(k'+1/2)^2(z-1)} \times \nonumber \\
	&\qquad \qquad \times 
	\frac{\partial}{\partial n_1}  \left[
	\sum\limits_{k=0}^{\infty} \frac{n_1}{(k+1/2)^2-n_1^2/n_2^2(k'+1/2)^2}e^{-\frac{2\pi^2}{n_1^2}(k+1/2)^2}
	\right] \Bigg\} \,.\label{eq:fig4}
	\end{align}
	This limit is illustrated in Fig.~4 of the main text, as a function of $z$.
\end{enumerate}

\section{Run-and-tumble particle}\label{appD}

\subsection{Derivation of Eq.~(13) in the main text} 
We consider the exit time distribution, $p_{\pm}(t,x)$, from the interval $[0,k]$,  for a run-and-tumble particle that starts at position $x$ with speed $\pm v$, respectively.  This distribution obeys the following coupled differential equations
\begin{align}
\begin{cases}
\partial _t p_+(t,x)= v \partial _x p_+(t,x)+\frac{1}{2T}\big[p_-(t,x)-p_+(t,x)\big]\\[2mm]
\partial _t p_-(t,x)= -v \partial _x p_-(t,x)+\frac{1}{2T}\big[p_+(t,x)-p_-(t,x)\big], \label{eq:PersisEq}
\end{cases}
\end{align}
with $ p_+(t=0,x)=p_-(t=0,x)=0$ and $p_+(t,k)=p_-(t,0)=\delta(t)$. Laplace transforming Eq.~\eqref{eq:PersisEq}, we obtain the Laplace transform of the exit-time distribution from an interval of length $k$ when the walk starts at position $dk$:
\begin{eqnarray}
\widehat{F}(s,k)&\equiv & \widehat{p}_+(s,dk)+\widehat{p}_-(s,dk) \nonumber \\
&=&\frac{(2T)^{-1}\big[\sinh(wdk)+\sinh(w(k-dk))+s\sinh(w(k-dk))+vw\cosh(w(k-dk))\big]}{s\sinh(wk)+vw\cosh(wk))+(2T)^{-1}\sinh(wk)} \; ,
\end{eqnarray}
where $w^2\equiv s^2/v^2+s/(v^2T)$. Keeping only the first-order terms in the limit $dk \ll k$, we have
\begin{eqnarray}
\widehat{F}(s,k)&=&1-wdk\;\frac{s\cosh(wk)+vw\sinh(wk)+(2T)^{-1}(\cosh(wk)-1)}{s\sinh(wk)+vw\cosh(wk)+(2T)^{-1}\sinh(wk)} \equiv 1+dk\, g(s,k)
\end{eqnarray}
with
\begin{eqnarray}
\int^0_n g(s,k)dk 
&=& \ln\left[\frac{1+\sqrt{\frac{s}{T^{-1}+s}} \tanh \left( wn/2 \right)}{1- \sqrt{\frac{s}{T^{-1}+s}} \tanh \left( wn/2 \right)}\frac{ sT+\cosh^2(wn/2)}{sT+1} \right]\equiv \ln(h(s,n)). \label{eq:persisth}
\end{eqnarray} 
To obtain the exit time distribution in the time domain from \eqref{eq:persisth} is difficult. However, in the diffusive limit $t \gg T$; i.e., $sT \ll 1$ and $n\sqrt{s}$ fixed, one can obtain simple results. By keeping the first order terms of $h$ in this limit, we get
\begin{eqnarray}
h(s,n)&\sim & \cosh ^2\left(\sqrt{\frac{s}{Tv^2}}\frac{n}{2}\right)\left[1+2\sqrt{sT}\tanh \left(\sqrt{\frac{s}{Tv^2}}\frac{n}{2}\right) \right]\,. \label{hcorrected}
\end{eqnarray}
Thus, with $D=Tv^2$, and defining  $\left \lbrace N_{T,v}(t)\right \rbrace $ as the span process for a run and tumble particle and $\left \lbrace N_{D}(t) \right\rbrace $ the span process for a Brownian motion, 
\begin{eqnarray}
\mathcal{L}\left( \mathbb{P}(N_{T,v}(t)\geq n)\right)=\mathcal{L}\left( \mathbb{P}(N_{D}(t)\geq n)\right) -\sqrt{\frac{4T}{s}}\frac{\sinh\left(\sqrt{ \frac{s}{D}}\frac{n}{2}\right)}{\cosh\left(\sqrt{ \frac{s}{D}}\frac{n}{2}\right)^3}+o(\sqrt{T/s}).
\end{eqnarray}
Consequently, we have that
\begin{eqnarray}
\mathbb{P}(N_{T,v}(t)\geq n)-\mathbb{P}(N_{D}(t)\geq n)&\sim&\partial^2_ n \left\{ \mathcal{L}^{-1}\left[ \frac{4D\sqrt{T}}{s^{3/2}}\tanh\left(\sqrt{ \frac{s}{D}}\frac{n}{2}\right)\right] \right\} \nonumber  \\
&\sim&- \partial^2_ n \left[  \sum\limits_{k=0}^{\infty}e^{-4D\pi ^2(k+1/2)^2t/n^2}\frac{4n\sqrt{DT}}{\pi ^2 (k+1/2)^2} \right] \,,
\end{eqnarray}
and differentiating this equation, we finally obtain
\begin{equation}
\mathbb{P}(N_{T,v}(t) = n)=\mathbb{P}(N_{D}(t)=n)+\partial^3_ n \left[\sum\limits_{k=0}^{\infty}e^{-4D\pi ^2(k+1/2)^2t/n^2}\frac{4n\sqrt{DT}}{\pi ^2 (k+1/2)^2} \right]+o(\sqrt{T}/t)\,.  
\end{equation}


\subsection{Comment on Eq.~(13) in the main text}

\subsubsection{First-order correction to the covariance}
We look at the $2$-time span distribution in the limit $  s_2 \ll s_1 \ll T^{-1}$  and $a_i=n_i\sqrt{s_i}$ fixed. We do the asymptotic development up to the first relevant order
\begin{subequations}
	\begin{align}
	&\mathcal{L}( \mathbb{P}(N_{T,v}(t_1)\geq n_1,N_{T,v}(t_2)\geq n_2) )  \nonumber \\
	&=\mathcal{L}( \mathbb{P}(N_{T,v}(t_1)\geq n_1) )\mathcal{L}( \mathbb{P}( N_{T,v}(t_2)\geq n_2) )\label{eq:indeppers}  \\
	&\quad+\frac{s_2}{s_1}\frac{ \left(a_1^2-2 a_1  \sqrt{D} \tanh \left(\frac{a_1}{2\sqrt{D}}\right)\right)}{4 D}  \mathcal{L}( \mathbb{P}(N_{D}(t_1)\geq n_1)  \mathbb{P}( N_{D}(t_2)\geq n_2) ) \label{eq:covRW} \\
	&\quad \hspace{-0.1cm}+\frac{\sqrt{T}}{s_1^{3/2}}\Bigg[ \frac{ -\left(a_1^2-2 a_1  \sqrt{D} \tanh \left(\frac{a_1}{2\sqrt{D}}\right)\right)\tanh\left(\frac{a_1}{2\sqrt{D}}\right)+D\left( \frac{a_1}{\sqrt{D}}+ \frac{a_1}{\sqrt{D}} \tanh ^2\left(\frac{a_1}{2\sqrt{D}}  \right)-2 \tanh \left(\frac{a_1}{2\sqrt{D}} \right)\right)}{2 D\cosh\left(\frac{a_1}{2\sqrt{D}}\right)^2\cosh ^2 \left(\frac{a_2}{2\sqrt{D}}\right)} \Bigg] \label{eq:partPERS} \\
	&\quad\hspace{-0.1cm} +\frac{T}{s_1} \frac{ \left(8 \left(a_1^2+3 D\right)-3 \left(3 a_1^2+8 D\right) \cosh^{-2}\left(\frac{a_1}{2 \sqrt{D}}\right)-\frac{a_1 \tanh \left(\frac{a_1}{2 \sqrt{D}}\right) \left(a_1^2-24 D \cosh^{-2}\left(\frac{a_1}{2 \sqrt{D}}\right)+34 D\right)}{\sqrt{D}}\right)}{8 D  \cosh^2 \left(\frac{a_1}{2\sqrt{D}}\right)\cosh^2 \left(\frac{a_2}{2\sqrt{D}}\right)}+o(T/s_1) \label{eq:relevantCOV}.
	\end{align}
\end{subequations}
Integration of Eq.~\eqref{eq:indeppers} gives the product of the first moments. Eq.~\eqref{eq:covRW} gives the same contribution to the covariance as the symmetric random walk. Integration of Eq.~(\ref{eq:partPERS}) with respect to $a_1$ and $a_2$ leads to the first-order correction to the difference of the covariances. However, this integral vanishes. This means that one should compute the correction at the next order, using \eqref{eq:relevantCOV}. This leads to
\begin{align}
& \text{Cov}\left(N_{T,v}(t_1), N_{T,v}(t_2) \right)-\text{Cov}\left(N_{D}(t_1), N_{D}(t_2) \right)\nonumber \\
&\sim \frac{2DT\sqrt{\frac{t_1}{ t_2}}}{\Gamma \left( 3/2\right) \Gamma \left( 1/2\right) }   \int_{0}^{\infty}da_1 \frac{ 8 \left(a_1^2+3 \right)-3 \left(3 a_1^2+8 \right) \cosh^{-2}\left(\frac{a_1}{2 }\right)-a_1 \tanh \left(\frac{a_1}{2 }\right) \left(a_1^2-24  \cosh^{-2}\left(\frac{a_1}{2 }\right)+34 \right)}{8   \cosh^2 \left(\frac{a_1}{2}\right)} \nonumber \\
&\approx -2.957 DT\sqrt{\frac{t_1}{t_2}}. \label{eq:CovFinal}
\end{align}

\subsubsection{First-order correction to the span distribution}
We study the correction to the diffusive limit of the two-time span distribution in the limit $T \ll t_1 \ll t_2$; i.e., $s_2T \ll s_1T \ll 1$. Keeping only the dominant terms, we have
\begin{align}
\mathcal{L}( \mathbb{P}(N_{T,v}(t_1)\geq n_1;N_{T,v}(t_2)\geq n_2) )&=\mathcal{L}( \mathbb{P}(N_{D}(t_1)\geq n_1;N_{D}(t_2)\geq n_2) ) \left[1 -2\sqrt{s_1T}\;\tanh\left(\sqrt{\frac{s_1}{D}}\frac{n_1}{2} \right) +o(\sqrt{s_1T})\right]\,.\label{eq:persist2time}
\end{align}
If we focus on the behavior of this distribution for a typical realization; i.e., situations for which $n_i\sqrt{s_i}=a_i$ is fixed, with $s_iT \to 0$ and $s_2/s_1 \to 0$, we have 
\begin{eqnarray}
\hspace{-0.5cm}\mathcal{L}( \mathbb{P}(N_{T,v}(t_1)\geq n_1;N_{T,v}(t_2)\geq n_2)-\mathcal{L}( \mathbb{P}(N_{D}(t_1)\geq n_1;N_{D}(t_2)\geq n_2) ) \propto \frac{\sqrt{T}}{s_1^{1/2}s_2}. 
\end{eqnarray}
This means that in the time domain the relative difference between the two cumulative distributions in the case where $n_i/\sqrt{t_i}=a_i$ is fixed behaves as $\propto \sqrt{T/t_1}$.

\section{Biased random walk}\label{appE}
\subsection{Derivation of Eq.~(14) in the main text}
First, we write the continuum limit exit-time distribution of the biased random walk of speed $v=p-q$, see Eq.~(2.2.28) of \cite{Redner:2001}, when the walk starts at distance $dk=1$ to the left boundary of the interval of size $k$ (and exits the interval either to the left or to the right side):
\begin{align}
\begin{split}
\widehat{F}^{l \to l}(s,k)=D \partial _x \left( \frac{e^{v(x-dk)/2D}}{Dw \sinh(wk) } \sinh(w(k-dk))\sinh(wx)\right)\bigg|_{x=0} = \frac{e^{-vdk/2D}}{ \sinh(wk) } \sinh(w(k-dk)) \, , \\[3mm]
\widehat{F}^{l \to r}(s,k)=-D \partial_x \left( \frac{e^{v(x-dk)/2D}}{Dw \sinh(wk) } \sinh(w(k-x))\sinh(wdk)\right)\bigg|_{x=k} = \frac{e^{v(k-dk)/2D}}{\sinh(wn) } \sinh(wdk) \, ,
\end{split}
\end{align}
with $w \equiv \sqrt{v^2+4Ds}/(2D)$.  Here the superscripts refer to the left and right ends of the interval. Similarly, by reversing the direction of the velocity, we have
\begin{align}
\begin{split}
\widehat{F}^{r \to r}(s,k)&= \frac{e^{vdk/2D}}{ \sinh(wk) } \sinh(w(k-dk)) \, , \\[2mm]
\widehat{F}^{r \to l}(s,k)&=\frac{e^{-v(k-dk)/2D}}{ \sinh(wk) } \sinh(wdk) \, .
\end{split}
\end{align}
We use the convention $\widehat{F}^{0 \to r}(s,0)=\widehat{F}^{0 \to l}(s,0)= 1/2$, corresponding to starting with a single visited site at time 0, which is both on the left and right side of the interval of size 1. We define the indices $\delta_i$ direction (left, $\delta_i=l$, or right, $\delta_i=r$, end of the visited interval) of the random walker when the site $i+1$ is first visited. Performing the same calculation as in Appendix \ref{appB} we get
\begin{align}
\nonumber
&\mathcal{L}\left( \mathbb{P}\left( N(t_1) \geqslant n_1;\ldots ; N(t_k) \geqslant n_k \right) \right) \nonumber \\
&\qquad=\frac{1}{s_1\ldots s_k}\sum\limits_{\lbrace \delta_i \rbrace_i}\widehat{F}^{0\to \delta_1}(s_1+\ldots +s_k,0)\ldots \widehat{F}^{\delta_{n_1-1}\to \delta_{n_1}}(s_1+\ldots +s_k,n_1-1) \nonumber \\
&\qquad\qquad\times \widehat{F}^{\delta_{n_1 }\to \delta_{n_1+1}}(s_2+\ldots +s_{k},n_1)\ldots \widehat{F}^{\delta_{n_k-1}\to \delta_{n_k}}(s_k,n_k-1) \, .
\end{align}

We use the transfer matrix technique by defining
\begin{eqnarray}
\widehat{F}(s,k) \equiv \begin{pmatrix}
\widehat{F}^{l \to l}(s,k) & \widehat{F}^{l \to r}(s,k)  \\
\widehat{F}^{r \to l}(s,k) & \widehat{F}^{r \to r}(s,k) \\
\end{pmatrix}.
\end{eqnarray}
In the limit $dk \ll  k$ (taking $w(s) k$ fixed), we obtain $\widehat{F}(s,k) = I_2+g(s,k)dk$, with
\begin{eqnarray}
g(s,k) &\equiv & \begin{pmatrix}
\frac{-v}{2D}-\frac{w}{\tanh(kw)} & w\frac{e^{vk/2D}}{\sinh(wk)}  \\
w\frac{e^{-vk/2D}}{\sinh(wk)} & \frac{v}{2D}-\frac{w}{\tanh(kw)} \\
\end{pmatrix}.
\end{eqnarray}
Using this matrix notation, we have the simpler formula 
\begin{align}
& \mathcal{L}\left( \mathbb{P}\left( N(t_1) \geqslant n_1;\ldots ; N(t_k) \geqslant n_k \right) \right) \nonumber \\
&\quad = \frac{1}{2s_1 s_2\ldots s_k}U^T\cdot \widehat{F}(s_1+\ldots +s_k,1) \cdot \ldots  \cdot \widehat{F}(s_1+\ldots +s_k,n_1-1) \cdot \widehat{F}(s_2+\ldots +s_k,n_1) \cdot \ldots  \cdot \widehat{F}(s_k,n_k-1) \cdot U \, . \nonumber \\
\end{align}
where the vector $U$ is $\big(\begin{smallmatrix}
1\\
1
\end{smallmatrix}\big)$.
Performing the product of matrices $\widehat{F}(s,k)$ having the same argument $s$, we rewrite the previous equation as
\begin{align}
&\mathcal{L}\left( \mathbb{P}\left( N(t_1) \geqslant n_1;\ldots ; N(t_k) \geqslant n_k \right) \right) \nonumber \\
&\quad = \frac{1}{2s_1 s_2\ldots s_k}\;U^T\cdot M(s_1+\ldots +s_k,0,n_1) \cdot M(s_2+\ldots +s_k,n_1,n_2) \cdot \ldots  \cdot M(s_k,n_{k-1},n_k) \cdot U \label{eq:ktimebiased}
\end{align}
where 
\begin{eqnarray}
M(s,m,n) \equiv~~  :\!\exp \left(\int_m^n g(s,k) dk \right)\!: \quad\mbox{ for $m<n$}, \label{eq:MatrixOrder}
\end{eqnarray}
and $:\ldots :$ represents the $k$-ordering operator. Eq.~\eqref{eq:MatrixOrder} is equivalent to the set of partial differential equations
$\partial_n M(s,m,n)=M(s,m,n)\cdot g(s,n)$ with $M(s,m,m)=I_2$.

\subsection{The one-time span distribution}
\begin{figure}
	\centering
	\includegraphics[scale=0.5]{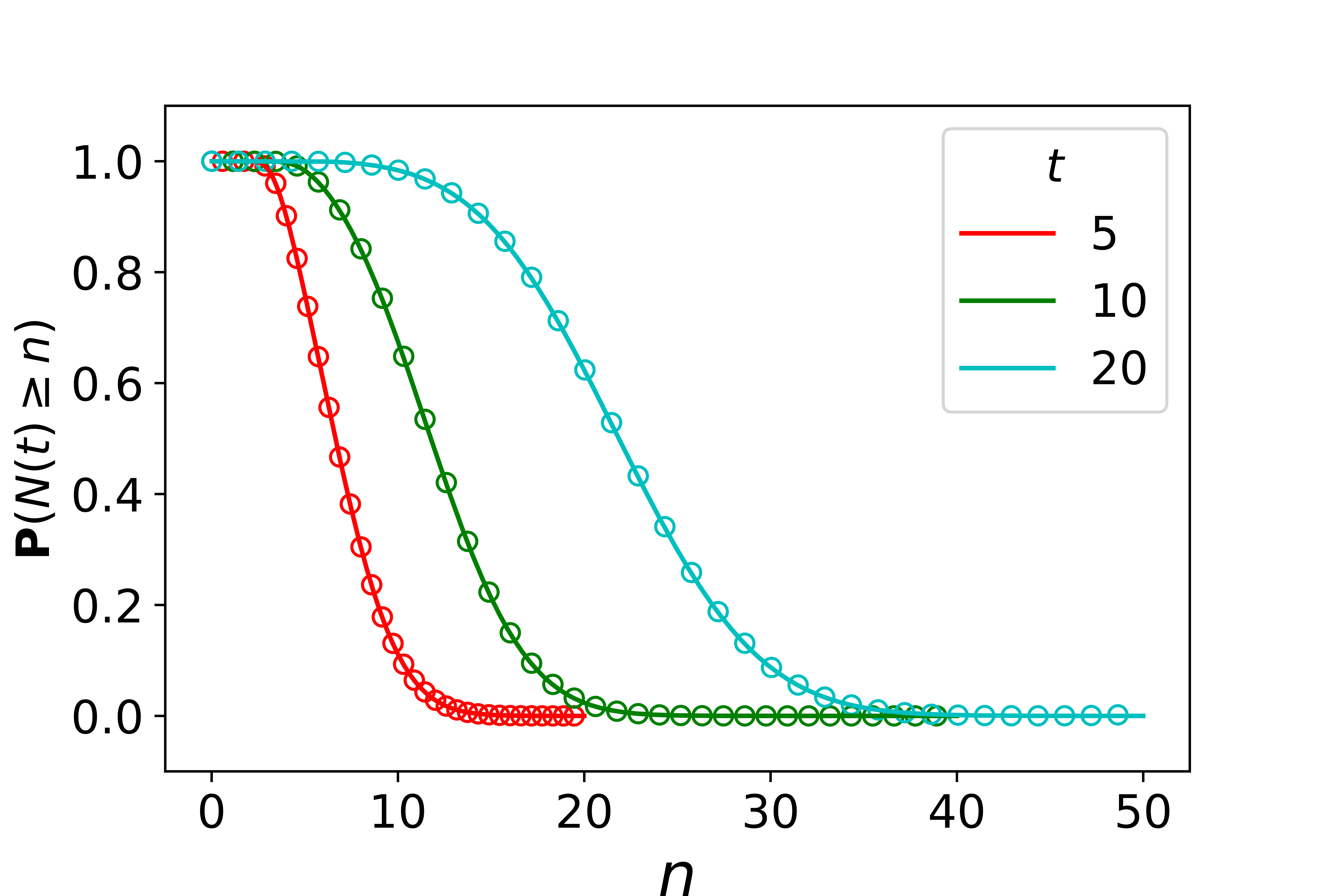}
	\caption{Cumulative distribution of the span at times $t=5,$ $10$ and $15$ for parameters $v=1$ and $D=1$.  The curves are $\frac{1}{2}(y_l(t,n)+y_r(t,n))$, while the circles are from \cite{wiese2020span}.}
	\label{fig:wiese}
\end{figure}

The one-time span distribution derived in \cite{wiese2020span} can be retrieved using Eq.~\eqref{eq:ktimebiased}. Consider the Laplace transform of the distribution of the number of distinct sites visited at a single time,
\begin{eqnarray}
\mathcal{L}\left(\mathbb{P}(N(t) \geq n \right)&=&\frac{1}{2s} U^T\cdot M(s,0,n) \cdot U=\frac{1}{2s}  x(s,n)^T \cdot U
\end{eqnarray}
where $x(s,n)$ is solution of $\partial_n x(s,n)= g(s,n)^T\cdot x(s,n)$ with $x(s,0)=U$ as follows from Eq.~\eqref{eq:MatrixOrder}.
We solve the system of equations for $n>0$ (for the sake of simplicity, we drop the argument $s$ and write $x(s,n)=(x_l(s,n),x_r(s,n))=(x_l(n),x_r(n))$):

\begin{equation}
\begin{cases}
\label{Biased1stStep}  
\partial_n x_l(n)=(-v/2D-w\coth(wn))x_l(n)+w\,\frac{\exp(-vn/2D)}{\sinh(wn)}\,x_r(n)\\[3mm]
\partial_n x_r(n)=(v/2D-w\coth(wn))x_r(n)+w\,\frac{\exp(vn/2D)}{\sinh(wn)}\,x_l(n) \; .
\end{cases}\
\end{equation}
Solving these coupled equations, we obtain
\begin{eqnarray}
x_r(n)&=&(v/2D-w \coth(wn))\;\frac{2}{\sinh(nw)}\int_0^n e^{v(n-n')/2D}\sinh(w n') dn' +2\,.
\end{eqnarray}
To obtain the distribution of the number of distinct sites visited in the time domain, we perform the inverse Laplace transform of $x_r(n)/s$, identifying the poles at $s=0$ and $w(s)n=ik\pi$, $k \in \mathbb{N}$:
\begin{align}
y_r(t,n) & \equiv \int_{-i\infty}^{+i\infty} \frac{dz}{2 \pi i} e^{zt} \frac{x_r(z,n)}{z} \\
&= \int_{-i\infty}^{+i\infty} \frac{dz}{2 \pi i} e^{zt} \frac{(v/2D-w(z) \coth(w(z)n))\;\frac{2}{\sinh(nw(z))}\int_0^n e^{v(n-n')/2D}\sinh(w(z) n') dn' +2}{z} \\
&=2+\left( \frac{v}{D\sinh(vn/2D)}-\frac{v\cosh(vn/2D)}{D\sinh(vn/2D)^2} \right)\int_0^{n} e^{v(n-n')/2D}\sinh(vn'/2D) dn'  \nonumber \\
&\qquad +\sum_{k=1}^{\infty} (-1)^k\int_0^{n} dn' e^{v(n-n')/2D} e^{-\frac{v^2t}{4D}} \left[ \frac{2k\pi  v \sin(k \pi n'/n) }{v^2n^2/4D+k^2\pi^2D}e^{-\frac{k^2\pi^2Dt}{n^2}}+\partial _n \left( \frac{4k\pi  D \sin(k \pi n'/n)}{v^2n^2/4D+k^2\pi^2D}e^{-\frac{k^2\pi^2Dt}{n^2}} \right) \right] \,. \label{eq:biasedright}
\end{align}
By reversing the velocity, $v \to -v$, we get the inverse Laplace transform of $x_l(n)/s$, namely, $y_l(t,n)$. We check numerically in \ref{fig:wiese} that $\mathbb{P}(N(t)\geq n) =\frac{1}{2}(y_l(t,n)+y_r(t,n))$ is indeed the same as the expression obtained in \cite{wiese2020span}. 

\subsection{Derivation of the $k$-time span distribution}
For the $k$-time span distribution, we need the general expression of the matrices $M_{m,n}(s)$. We solve the set of partial differential equations \eqref{Biased1stStep} starting at $m>0$ to $n>m$ with arbitrary initial conditions $(x_l(m),x_r(m))$. 
Defining $\tilde{x}_l(n)\equiv x_l(n)\sinh(wn) \exp(vn/2D)$ and $\tilde{x}_r(n)\equiv x_r(n)\sinh(wn) \exp(-vn/2D)$, we have that
\begin{equation}
\begin{cases}
\partial_n \tilde{x}_l(n)=w x_r(n) \\
\partial_n \tilde{x}_r(n)=w x_l(n) \; .
\end{cases}\
\end{equation}
Thus, as $\tilde{x}_l(m)=x_l(m)\exp(vm/2D)\sinh(wm)$ and $\tilde{x}_r(m)=x_r(m)\exp(-vm/2D)\sinh(wm) $,
\begin{align}
\begin{cases}
x_l(n)=w \frac{\exp(-vn/2D)}{\sinh(wn)} \left(\int_m^n x_r(n')dn' +w^{-1}x_l(m)\exp(vm/2D)\sinh(wm) \right)=w \frac{\exp(-vn/2D)}{\sinh(wn)}X_r(n) \\[3mm]
x_r(n)=w \frac{\exp(vn/2D)}{\sinh(wn)} \left( \int_m^n x_l(n')dn'+w^{-1}x_r(m)\exp(-vm/2D)\sinh(wm) \right)=w \frac{\exp(vn/2D)}{\sinh(wn)}X_l(n) \; .
\end{cases}\
\end{align}
This gives the equation for $X_r$
\begin{eqnarray}
\partial_n X_r(n)&=&(v/2D-w\coth(wn))X_r(n)-(v/2D-w\coth(wm))X_r(m)+x_r(m) \\
&=&(v/2D-w\coth(wn))X_r(n)+\alpha(m) \; , \label{eq:Intermed}
\end{eqnarray}
where $\alpha(m)$ is defined as 
\begin{eqnarray}
\alpha(m)&\equiv &w^{-1}(-v/2D+w\coth(wm))x_l(m)\sinh(w m)\exp(vm/2D)+x_r(m) \; .
\end{eqnarray}
Eq. \eqref{eq:Intermed} solution is given by
\begin{eqnarray}
X_r(n)&=&\alpha(m)\frac{\exp(vn/2D)}{\sinh(wn)}\int _m^n \exp(-vn'/2D)\sinh(wn')+X_r(m)\frac{\exp(v(n-m)/2D)\sinh(wm)}{\sinh(wn)} 
\end{eqnarray}
and deriving this expression,
\begin{eqnarray}
x_r(n)&=& (v/2D-w\coth(wn)) \Bigg(\alpha(m)\frac{\exp(vn/2D)}{\sinh(wn)}\int _m^n \exp(-vn'/2D)\sinh(wn') \nonumber \\
&&+X_r(m)\frac{\exp(v(n-m)/2D)\sinh(wm)}{\sinh(wn)} \Bigg)+\alpha (m).
\end{eqnarray}
We get the expression for $x_r(n)$ as a function of $x_l(m)$ and $x_r(m)$
\begin{align}
x_r(n)&=x_r(m)\left( (v/2D-w\coth(wn))\frac{\exp(vn/2D)}{\sinh(wn)}\int _m^n \exp(-vn'/2D)\sinh(wn')dn' +1 \right) \nonumber \\
&+x_l(m) \Bigg[- w^{-1}(v/2D-w\coth(wn))(v/2D-w\coth(wm))\frac{\exp(v(m+n)/2D)\sinh(wm)}{\sinh(wn)}  \nonumber \\
& \times \int _m^n \exp(-vn'/2D)\sinh(wn')dn' -w^{-1}(v/2D-w\coth(wm))(\exp(vm/2D)\sinh(wm) \nonumber \\
&+w^{-1}(v/2D-w\coth(wn))(\exp(vm/2D)\sinh(wm) \exp(-v(n-m)/2D) \frac{\sinh(wm)}{\sinh(wn)} \Bigg] \,. \label{eq:xr}
\end{align}
The expression for $x_l(n)$ is similar and is obtained by reversing the velocity $v \to -v$ and the indices $l \leftrightarrow r$.
Thus we obtain, using (\ref{eq:xr}), the exact expression of the matrix $M(s,m,n)$ 
\begin{eqnarray}
M(s,m,n) \equiv \begin{pmatrix}
\mu(-v,m,n) & \nu(v,m,n)  \\
\nu(-v,m,n) & \mu(v,m,n) \\
\end{pmatrix}, \label{eq:matrix}
\end{eqnarray}
with
\begin{eqnarray}
\mu(v,m,n)&\equiv& 1+(v/2D-w \coth(wn))\left(\frac{\exp(vn/2D)}{\sinh(wn)}\int_m^n\exp(-vn'/2D)\sinh(w n')dn' \right) \label{eq:mu} \\
\nu(v,m,n) &\equiv& \frac{e^{vm/2D}\sinh(wm)}{w}\Bigg[(v/2D-w\coth(wn))\Bigg(-\frac{v/2D-w\coth(wm)}{\sinh(wn)}\int_m^n e^{\frac{v(n-n')}{2D}}\sinh(wn')dn'  \nonumber \\
&&  +\frac{\exp(v(n-m)/2D)\sinh(wm)}{\sinh(wn)}\Bigg)-(v/2D-w\coth(wm)) \Bigg]\,, \label{eq:nu}
\end{eqnarray}
and all the $s$ dependence being within the variable $w$.
Hence, the knowledge of Eqs.~(\ref{eq:ktimebiased}), (\ref{eq:matrix}), (\ref{eq:mu}), (\ref{eq:nu}) give us the full description of the stochastic process $\left \lbrace N(t) \right \rbrace $, the span of a biased random walk.

\section{Illustration of Eq. (13) and (14)}\label{appF}
\begin{figure}[ht!]
	\includegraphics[width=0.5\columnwidth]{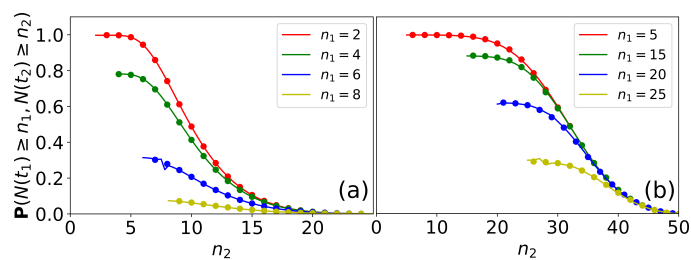}
	\caption{\label{PersistBiased}$2$-time distribution for the persistent random walk for parameters $T=v=1$ and  $t_1,t_2=10,30$ (a), as well as for a biased random walk of parameters $v=1$ and diffusion $D=1$, at times $t_1,t_2=20,30$ (b). Numerical simulations are represented by the dots.}
\end{figure}
We numerically invert the Laplace transform of the $2-$time distributions of a run-and-tumble particle using Eq.~(13) and of a biased random walk using Eq.~(14) of the main text. As shown in \ref{PersistBiased}, knowledge of the Laplace transformed distributions can be used to extract numerical values of the statistics, besides the theoretical characterization.  

\twocolumngrid

\bibliographystyle{apsrev4-1}
%

\end{document}